\documentclass[superscriptaddress,groupedaddress,nofootnoteinbib,11pt]{article}
\pdfoutput=1 % if your are submitting a pdflatex (i.e. if you have images in pdf, png or jpg format)
\usepackage{jheppub}

\usepackage{CJKutf8} % for Chinese characters
\usepackage{mathtools,slashed,mathrsfs}
\usepackage[caption=false]{subfig}
\usepackage{dcolumn}% Align table columns on decimal point
\usepackage{multirow}
\usepackage{tabularx}
\usepackage{booktabs}
\usepackage{bm}% bold math
\usepackage{amsmath}
\usepackage{comment}
\usepackage{setspace}
\usepackage[dvipsnames]{xcolor}
\usepackage[normalem]{ulem} % to strike through
\usepackage{enumerate}
\usepackage{siunitx}
\usepackage{float}
\usepackage{hyperref}

\newcommand{\MP}{M_P}

\definecolor{mypurple}{RGB}{164,64,214}

\newcommand\eea{\end{eqnarray}}
\newcommand\bea{\begin{eqnarray}}

\newcommand{\mpl}{M_{pl}}
\newcommand{\trh}{T_{RH}}

\def\l{\left(}
\def\r{\right)}

\newcommand\bes{\begin{split}}
\newcommand\ees{\end{split}}

\begin{document}

\title{Black Hole Production of Monopoles in the Early Universe}

\date{\today}
\author[a]{Saurav Das}
\author[a]{and Anson Hook}

\affiliation[a]{Maryland Center for Fundamental Physics, University of Maryland, College Park, MD 20742}
% e-mail addresses: one for each author, in the same order as the authors
\emailAdd{sauutsab@umd.edu}
\emailAdd{hook@umd.edu}

\abstract{
In the early universe, evaporating black holes heat up the surrounding plasma and create a temperature profile around the black hole that can be more important than the black hole itself.  As an example, we demonstrate how the hot plasma surrounding evaporating black holes can efficiently produce monopoles via the Kibble-Zurek mechanism. In the case where black holes reheat the universe, reheat temperatures above $\sim 500$ GeV can already lead to monopoles overclosing the universe.
%In the early universe, evaporating black holes can easily produce monopoles.  Via Hawking radiation, black holes heat up the surrounding plasma and create a temperature profile around the black hole that features symmetry restoration near the center.  Eventually this region cools off and undergoes the Kibble-Zurek mechanism, producing monopoles.  We demonstrate that this process can very efficiently produce monopoles.  In the case where black holes reheat the universe, reheat temperatures above $\sim 500$ GeV can already lead to monopoles overclosing the universe.
}

\maketitle

\section{Introduction}

One of the most fascinating objects in physics are black holes. 
Black holes have been observed at the center of many galaxies and play a central role in astrophysics~\cite{LIGOScientific:2016aoc,EventHorizonTelescope:2019dse}.
Aside from astrophysical black holes, black holes are often also a consequence of early universe dynamics such as hybrid inflation~\cite{Linde:1993cn,Garcia-Bellido:1996mdl,Garcia-Bellido:1997hex,lyth2011primordial,Bugaev:2011wy,Clesse:2015wea,Kawasaki:2015ppx,Kawasaki:2016pql}, phase transitions~\cite{Crawford:1982yz,PhysRevD.26.2681,Freivogel:2007fx,Moss:1994pi,Johnson:2011wt,Jedamzik:1999am,Baker:2021nyl,Lewicki:2019gmv} or topological defects~\cite{Hawking:1987bn,Polnarev:1988dh}.
In fact, in many of these scenarios there are so many black holes produced that they become the dominant energy density of the entire universe~\cite{Hawking:1971ei,Chapline:1975ojl,Carr:2016drx,Carr:2020xqk,Green:2020jor,Villanueva-Domingo:2021spv,Young:2015kda}.

While a black hole dominated era may seem like a phenomenological disaster, it turns out that due to Hawking radiation~\cite{Hawking:1974rv,Hawking:1975vcx,Page:1976df}, black holes evaporate and the universe eventually transitions into the standard early universe radiation dominated regime~\cite{Green:1999yh,Carr:2009jm,Lennon:2017tqq,Hooper:2019gtx,Hook:2014mla,Baldes:2020nuv,Keith:2020jww}.
As compared to more standard reheating mechanisms, reheating the universe through black hole evaporation is a rather unique process.
The reason for this is two-fold.  

Firstly, as a black hole evaporates, its temperature rises until the black hole becomes a Planck mass and Planck temperature object.  As such, black hole evaporation depends in part on ultraviolet physics.
Secondly, a black hole is extremely massive and thus serves as a heat source.  It heats up the surrounding plasma to large temperatures creating a temperature profile around the black hole that can also reach temperatures close to the Planck scale~\footnote{Even if there was originally no radiation around the black hole, once $\mathcal{O}(1)$ of the black holes have started to evaporate, the universe is effectively reheated and there is a large bath of particles for the rest of the black holes to heat up.}.  Due to the large radius of this profile, the effects of the surrounding plasma may be more significant than the black hole itself.

In this article, we present one example, monopole production, where the the hot plasma surrounding a black hole is more important than the black hole itself.  Monopoles are another object of great interest to particle physicists~\cite{Dirac:1931kp,Guth:1979bh,Preskill:1984gd}.  They are especially relevant given that they are a generic prediction of quantum field theories and feature in many well motivated models such as Grand Unified Theories~\cite{Georgi:1974sy,Georgi:1974yf,tHooft:1974kcl,Polyakov:1974ek}.
Aside from their magnetic charge, the other property of monopoles is that as composite objects, their physical radius is larger than their Compton wavelength.  This mismatch means that monopole production coming from Hawking radiation is always exponentially suppressed~\cite{Johnson:2018gjr}.  By the time that black holes are hot enough to produce monopoles without a Boltzmann suppression, the black hole is smaller than the monopole meaning that the emission of monopoles is still exponentially suppressed.

However, monopole production by the plasma surrounding the black hole is not exponentially suppressed.  Because the plasma surrounding the black hole has a radius much larger than the black hole, it can easily produce many monopoles via the Kibble-Zurek mechanism \cite{Kibble:1976sj,Zurek:1985qw,Murayama:2009nj}.  Close to the black hole, the plasma is hot enough that symmetry is restored.  After the black hole evaporates, this hot region slowly cools down.  At some point, it undergoes a symmetry breaking phase transition.  Regions of space separated by more than a correlation length all choose their vacua independently and monopoles are created by accident.

Monopole production from evaporating black holes can be extremely efficient.
As an example, we show that if the universe was reheated by black holes and monopoles were produced by a second order phase transition, the monopole over-production limits the reheat temperature to be
\bea
T_{RH} \lesssim 500 \, \text{GeV} \, \left ( \frac{10^{15} \, \text{GeV}}{T_{PT}} \right )^{9/35}.
\eea
Despite the very low reheat temperature of the universe and the very large scale associated with the phase transition, $T_{PT}$, producing the monopole, monopoles can still very easily over-close the universe.  
The re-introduction of the monopole problem in this context occurs because even if the average temperature is low, the temperature around the black holes themselves is still very large.

A reheat temperature this low is very impactful as it strongly favors reheat temperatures lower than the scale at which electroweak sphalerons are active.  This limits the available baryogenesis mechanisms and pushes one to consider black hole assisted baryogenesis mechanisms~\cite{Hawking:1974rv,Carr:1976zz,Turner:1979bt,PhysRevD.43.984,Majumdar:1995yr,Upadhyay:1999vk,Baumann:2007yr,Hook:2014mla,Fujita:2014hha,Hamada:2016jnq,Hooper:2020otu}.  Additionally, many models of black hole production in the early universe produce only a sub-population of black holes, e.g. a popular scenario is when primordial black holes are dark matter~\cite{Sasaki:2018dmp,Carr:2020gox,Green:2020jor}.  Evaporation of a sub-population of primordial black holes provides a mechanism for producing a sub-dominant populations of monopoles.

In Sec.~\ref{Sec: temp}, we derive the temperature profile of the plasma surrounding a black hole in the early universe.  
In Sec.~\ref{Sec: monopoles}, we calculate how many monopoles are produced per black hole.
In Sec.~\ref{Sec: reheat}, we place a bound on the reheat temperature coming from monopole over production if the universe was reheated by black holes.  
In Sec.~\ref{Sec: approx}, we discuss the approximations under which our calculations are valid.
Finally, we conclude in Sec.~\ref{Sec: conclusion}.

\section{Temperature Profile Around a Black Hole} \label{Sec: temp}

In this section, we derive the temperature profile around a black hole evaporating in the early universe.  We will describe how a black hole heats up the plasma surrounding it and how this hot region of space cools after the black hole evaporates.

\subsection{Radiation Transfer}\label{Sec: radtrans}

To set the stage, we first present the derivation of the equations governing the transfer of energy in a relativistic thermal system.  The discussion in this section will be a terse summary of the material presented in Ref.~\cite{Rybicki:2004hfl}.  

The starting point is a quantity called the specific intensity, $I_\nu(\vec r, \hat s , t)$, which is the energy emitted per area per solid angle per time per frequency.  For simplicity, we will work in the plane approximation where quantities vary in z but not in x and y.  
The specific intensity is related to the more familiar quantities such as a systems total energy density ($\rho$), radiation pressure ($P$), and energy flux ($J_z$) by
\bea
\rho = \int d\Omega d\nu \, I_\nu \qquad 
J_z = \int d\Omega d\nu \cos \theta \, I_\nu \qquad 
P = \int d\Omega d\nu \cos^2 \theta \, I_\nu. \nonumber 
\eea

The energy emitted in a direction $\hat s$ has its power modulated by
\bea
\label{Eq: rad1}
\frac{dI_\nu}{ds} = - \frac{1}{\lambda} I_\nu + j_\nu
\eea
where $j_\nu$ is the power being emitted in the $\hat s$ direction and $\lambda = 1/n \sigma$  is the scattering length.
The first term gives the absorption of radiation passing through the thermal bath while the second gives the power emitted by the bath itself.
Often, Eq.~\ref{Eq: rad1} is written in terms of the optical depth parameter by $d \tau = -dz/\lambda = - \cos \theta \, ds/\lambda$.  Written in terms of the optical depth, Eq.~\ref{Eq: rad1} becomes
\bea
\cos \theta \frac{d I_\nu}{d \tau} = I_\nu - \lambda j_\nu.
\eea
This equation can be further simplified by multiplying by $\cos \theta$ and integrating over solid angle and frequency.  
A thermal system emits radiation isotropically so that $\int d\Omega \cos \theta \, j_\nu = 0$, leaving
\bea
\label{Eq: flux}
J_z = \frac{dP}{d\tau} = - \frac{\lambda}{3} \frac{d \rho}{d z},
\eea
where in the second equality we used the fact that in thermal equilibrium $P = \rho/3$.  Eq.~\ref{Eq: flux} shows how the energy flux is related to temperature gradients.
The final equation we will consider is simply the conservation of energy
\bea
\label{Eq: solve}
\frac{d \rho}{dt } = - \frac{d J_z}{d z} = \vec \nabla \left ( \frac{\lambda}{3} \vec \nabla \rho \right ) ,
\eea
where we use Eq.~\ref{Eq: flux} in the second equality. 
% \begin{equation}\label{eq:diffusion}
%     \frac{\partial T^4(r,t)}{\partial t}=\frac{\pi^2}{3\zeta(3)c_{\alpha}^2g_*}\frac{1}{r^2}\frac{\partial }{\partial r}\left(\frac{r^2}{T(r,t)}\frac{\partial T^4(r,t)}{\partial r}\right)
% \end{equation}
This conservation equation governs how a system heats and cools.
In what follows we will be solving this conservation equation with various initial conditions and boundary conditions corresponding to a black hole evaporating and the subsequent cooling phase.  In a relativistic thermal system $\lambda \propto 1/T$ and $\rho \propto T^4$, so Eq.~\ref{Eq: solve} is a differential equation that one can solve for the temperature profile, $T(\vec r,t)$.

\subsection{Heating} \label{Sec: heat}

In this subsection, we will solve Eq.~\ref{Eq: solve} subject to the condition that there is a black hole providing a heat source at $r= 0$.
In the cases we will be interested in, we are dealing with regions close enough to the black hole that we can safely assume that the asymptotic temperature is negligible.

Intuition for the problem can be built up by first starting with the case of a boundary condition $T = T_0$ at a radius $r_0$.  In this case, the equilibrium solution of Eq.~\ref{Eq: solve} can be easily seen to be
\bea
\label{Eq: EQprofile}
T(r) = \left( \frac{r_0}{r} \right )^{1/3} T_0.
\eea
Thus we see that when in equilibrium, the temperature falls off rather slowly when far from the heat source.
The above scaling can also be obtained by using the fact that in equilibrium, the total energy leaving any radius must be the same.  Using Eq.~\ref{Eq: flux} we have $4 \pi r^2 J_r \sim r T^3$, which should be radius independent, giving another way of finding Eq.~\ref{Eq: EQprofile}.

We now discuss the situation of interest.  A black hole of mass $M_{BH}$ initially starts off emitting Hawking radiation at a temperature $T_{BH}$ from a radius $r_{BH}$,
\bea
T_{BH} = \frac{\mpl^2}{8 \pi M_{BH}} \qquad r_{BH} = \frac{1}{4 \pi T_{BH}},
\eea
where $\mpl$ is Planck's constant.  The Hawking radiation receives grey body correction factors so that the total energy emitted by the black hole is
\bea
\frac{d M_{BH}}{dt} = - \frac{\mathcal{G}_f g_{\star} \mpl^4}{30720 \pi M_{BH}^2} \equiv -c_1 \frac{\mpl^4}{M_{BH}^2},
\eea
where $\mathcal{G}_f$ is the temperature and spin dependent grey body factor and $g_*(T)$ is the total number of entropic degrees of freedom.  For the Standard Model,  we average over all particles and find that $\mathcal{G}_f \sim 4$.  As a result of this emission, the black hole slowly evaporates and its temperature as a function of time is
\bea
\label{Eq: tbh of t}
T_{BH}(t) = \frac{\mpl^2}{8 \pi} \frac{1}{\left ( M_{BH}^3 - 3 c_1 \mpl^4 t \right )^{1/3}}.
\eea
In the rest of this sub-section, we will derive the temperature profile that results from a time and space dependent boundary condition of this sort.

\paragraph{Analytical Estimate}\label{Sec:anaheating}

We first describe how to obtain an $\mathcal{O}(1)$ analytical estimate of the profile before moving on to numerical solutions.  
There are two critical observations that render black hole heating easy to estimate.
The first fact is that in any thermal system, energy diffuses out as a random walk.
If there is a change at $r=0$ at a time $t=0$, then at a later time $t$ only radii smaller than $r^2 \lesssim \lambda t$ have noticed the change while all physics outside of this radius have not noticed the change.
The second observation is the simple fact that when the black hole evaporates, it spends more time at lower temperatures than it does at higher temperatures.
Combined, these two observations allow one to treat the heating process like an onion.  At any given time, the black hole is in diffusion limited thermal contact with a region of space around it.  As the black hole evaporates, this region of space shrinks.  The final temperature profile at any given radius is set by when that radius ``freezes-out" and leaves equilibrium with the black hole.

To turn this intuition into a set of equations, we start with a black hole of temperature $T_{BH}$.  This black hole evaporates in a finite time $t_{BH} \sim \mpl^2/T_{BH}^3$.  In this time, the energy can only diffuse out to a distance
\bea
\label{Eq: diffusion}
r_d^2 \sim \lambda t_{BH} \sim \frac{\mpl^2}{T(r_d) T_{BH}^3}.
\eea
We will make the approximation that everything inside of $r_d$ is in equilibrium with the black hole while everything outside of $r_d$ is not in equilibrium and has been frozen in place and is no longer changing.  
When we later consider cooling, it will become clear that the cooling time is much longer than the heating time and that treating the temperature profile outside of the radius $r_d$ as constant is a reasonable approximation.

A consequence of this freezing-out assumption is that for radii $r < r_d$, the temperature profile should scale as $T \sim 1/r^{1/3}$ as found in Eq.~\ref{Eq: EQprofile}.  While for radii $r > r_d$, the temperature profile is a fossilized memory of when the black hole was just leaving equilibrium with that radius.  Eventually the black hole has evaporated completely and the temperature profile consists entirely of the frozen out part of the profile.

To determine the final temperature profile after the evaporation of the black hole, at every radius, we find what its temperature was when it was just leaving equilibrium with the black hole and set the final temperature to be that value.
In equilibrium, the flux of energy leaving the black hole is equal to the flux leaving the thermal bath.  Applying this to the radius $r_d$, we have
\bea
\label{Eq: radiate energy}
\frac{d M_{BH}}{dt} \sim T_{BH}^2 \sim 4 \pi r_d^2 J_r \sim r_d T(r_d)^3.
\eea
Combining Eq.~\ref{Eq: diffusion} and Eq.~\ref{Eq: radiate energy} we arrive at the scaling
\bea
\label{Eq: heat}
T(r) \sim \frac{\mpl^{4/11}}{r^{7/11}}
\eea
for all radii which have decoupled from the black hole.

There is a second derivation of this scaling where instead of requiring Eq.~\ref{Eq: radiate energy}, we instead impose that the black hole's mass is larger than the total energy stored in the region $r_d$.  If the black hole mass is smaller, then that temperature is necessarily frozen out by conservation of energy.  This different argument also gives the scaling found in Eq.~\ref{Eq: heat}.

\paragraph{Numerical calculation}

To put $\mathcal{O}(1)$ numbers to Eq.~\ref{Eq: heat}, we will solve Eq.~\ref{Eq: solve} numerically.
Unfortunately, the exact problem of interest has two aspects which makes it difficult to solve numerically. 
The first is that the black hole shrinks as it evolves and a boundary condition whose location $r_{H}$ is changing as a function of time is difficult to solve. 
The second one is that the region right near the black hole is not in thermal equilibrium with the radiation emitted by the black hole and thus would require other methods to deal with.\\
% This problem lets us numerically find the $\mathcal{O}(1)$ number associated with the approximation used in Eq.~\ref{Eq: diffusion}. 

\begin{figure}[t]
    \centering
    \includegraphics[width=0.7 \textwidth]{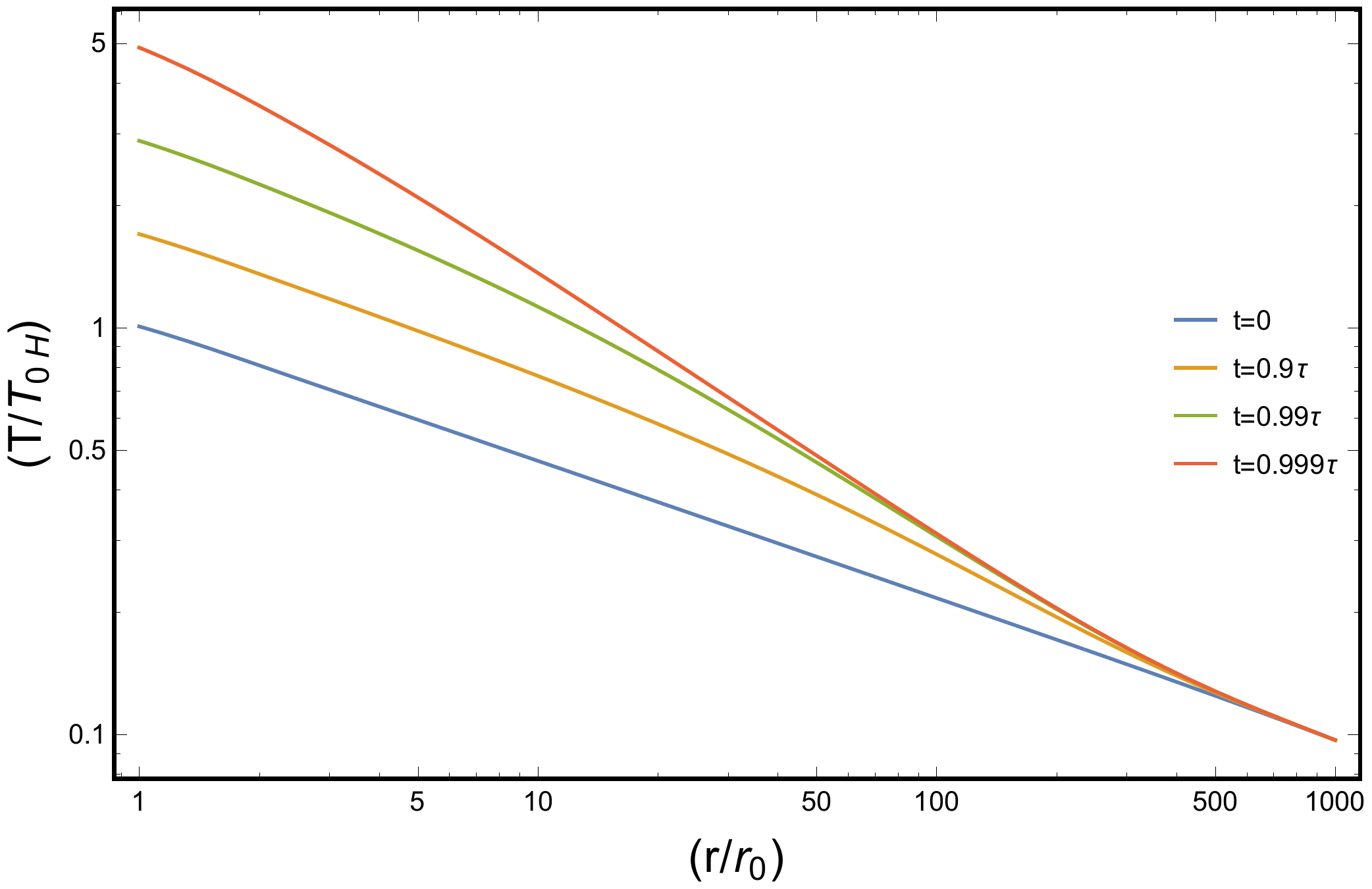}
    \caption{The temperature profile during the heating phase of a black hole like object.  We start from an equilibrium profile at $t=0$ of $T(r)\sim r^{-1/3}$. The temperature profile below the diffusion radius remains in equilibrium with the black hole. As the diffusion radius gradually shrinks, temperatures outside of it freezes out giving a profile of $T(r)\sim r^{-7/11}$ as expected from analytical arguments of this example. }
    \label{fig:my_label}
\end{figure}

To eliminate these problems, we consider a fixed radius, $r_0$ which is in thermal contact with the black hole and use it as our boundary.  Any surface which is in thermal contact with the black hole has the same energy passing through it as what was emitted by the black hole. This means 
\bea
\label{Eq: boundary1}
&\frac{d M_{BH}}{dt} \sim T_{BH}^2 \sim 4 \pi r_0^2 J_{r_0} \sim r_0 T(r_0)^3\\ 
&\implies T(r_0)\sim \frac{T_{BH}^{2/3}}{r_0^{1/3}}.
\eea
%We further assume that thermal equilibrium is maintained throughout black hole evaporation to remedy the second problem. This assumption breaks down at the very last stages of the evaporation and our solution can be trusted outside that regime. 
We can then solve Eq.~\ref{Eq: solve} with the boundary condition
\bea
\label{Eq: boundary condition}
T(r_0,t)\equiv T_0(t)=T_{0H}\left(\frac{\tau}{\tau-t}\right)^{\frac{2}{9}},
\eea
which was obtained from Eq.~\ref{Eq: boundary1} using Eq.~\ref{Eq: tbh of t}.
$\tau$ is the characteristics lifetime of the black hole and $T_{0H}$ is the initial temperature of the surface at $r_0$.  To match our numerical solution to the analytical approach, we repeat the analysis done in the analytic section including a proportionality coefficient that will be determined numerically.

Energy passing through the surface at $r_0$ can only reach a finite distance during the lifetime of the black hole.  The lifetime of the black hole as a function of the temperature at $r_0$ is  
% where $\tau$ is some characteristics lifetime of the system and $T_{0H}$ is the initial temperature.  Similar to the black hole, the radiated energy can only reach a finite distance in its lifetime and different radii freeze-out at different times. Although the object never evaporates, its effective lifetime as a function of its temperature is 
\bea
t_c = \tau \l\frac{T_{0H}}{T_0}\r^{9/2} .
\eea
Within its lifetime, the energy can only diffuse up to a distance $r_d$
\bea
\label{Eq: num1}
r_d^2=c_0 \lambda \, t_c = c_0 \frac{c_\lambda}{T(r_d)}\tau \l\frac{T_{0H}}{T_0}\r^{9/2},
\eea
where we have introduced a proportionality constant $c_0$ that will be determined numerically.  We have also defined $c_\lambda$ as $\lambda(T)=1/(n\sigma)=c_\lambda/T$, which is a parameter that depends on the microscopics of the theory and can be scaled out of the problem. \\
As before, we require that the flux emitted at $r_0$ is the same as the flux passing through $r_d$,
\bea
\label{Eq: num2}
r_0 T_0^3 = r_d T(r_d)^3.
\eea
Combining Eq.~\ref{Eq: num1} and Eq.~\ref{Eq: num2}, we find the scaling
\bea
\label{Eq: scaling}
T(r)=\left(c_0 c_\lambda \right)^{2/11}  \frac{\l r_0 T_{0H}^3\r^{3/11}\tau^{2/11}}{r^{7/11}} . 
\eea
We have solved Eq.~\ref{Eq: solve} numerically subject to the boundary condition shown in Eq.~\ref{Eq: boundary condition}.  Perhaps unsurprisingly, the numerical solution exhibits the scaling found in Eq.~\ref{Eq: scaling} with 
\bea
c_0=0.59.
\eea

The numerical solution is shown in Fig.~\ref{fig:my_label} for various times.  For simplicity, an initial condition of $T \sim 1/r^{1/3}$ is assumed.  Many of the features anticipated by the analytical analysis are found here.  First, all radii beyond $r \sim 700 r_0$ are frozen to their initial values.  The black hole evaporates too quickly for the heat it deposits to diffuse past that radius.  The second feature that is present at all times and most visible for $t = 0.9 \tau$, is that near the black hole there is a region of space which is still in equilibrium with the black hole and scaling as $T \sim 1/r^{1/3}$.  The last visible feature, most easily seen for $t \gtrsim 0.99 \tau$,  is the freeze-out regime.  The regions beyond $r \gtrsim 50 \, r_0$ have frozen out and subsequent evolution is not changing its temperature profile.

The constant $c_0$ allows us to calculate
Eq.~\ref{Eq: heat} for the problem of interest with $\mathcal{O}(1)$ numbers, namely the heating profile of an evaporating black hole. To facilitate our description of the calculation, we define new constants $c_2$ and $c_5$ as 
\begin{equation}
    \begin{split}
        %\frac{dM}{dt}&=-c_1\frac{\mpl^4}{M^2}\\ 
        t_{BH}\equiv c_2\frac{\mpl^2}{T_{H}^3} \qquad \qquad 
        \rho\equiv c_5T^4 .
    \end{split}
\end{equation}
We can now re-derive Eq.~\ref{Eq: heat} in all of its full numerical glory
\begin{equation}\label{Eq:heatingprofile}
    T(r)=\l\frac{6^6\pi^3c_0^2c_1^2c_2^2}{c_5^3c_4}\r^{1/11}\frac{\mpl^{4/11}}{r^{7/11}}= 0.183 \l{\frac{c_{\alpha} ^2 G_f}{g_*(T)}}\r^{1/11}\frac{\mpl^{4/11}}{r^{7/11}}
\end{equation}
Where as before $G_f$ is the Grey body factor of a black hole, $g_*(T)$ is the total number of entropic degrees of freedom and $c_{\alpha}$ characterizes the scattering cross section, $\sigma(T)=c_{\alpha}^2(T)/T^2$.  The expectation is that $c_{\alpha}$ will be of order $\alpha = g^2/4 \pi$, but due to the large number of possible final states, $c_\alpha$ can be a bit larger than $\alpha$.
The main assumption we have made so far is that is thermal equilibrium is maintained.  Eventually the black hole's Hawking radiation will not be instantly absorbed by the thermal bath so that Eq.~\ref{Eq:heatingprofile} is only valid for distances larger than some critical radius.

\subsection{Cooling}

Right after the black hole has evaporated, the temperature profile around it is of the form $T(r) \equiv c \, \mpl^{4/11}/r^{7/11}$.  In this subsection, we describe how this temperature profile cools.

\paragraph{Analytical Estimate}

The temperature profile left after a black hole evaporates is IR dominated so that it cools through an inverse of how it heated.  Namely, the inner regions cool faster than the outer regions.  To see this explicitly, we can use Eq.~\ref{Eq: solve} to see that  
\bea
t_{char} \sim \frac{E}{dE/dt} \sim \frac{r^3 T^4}{\frac{r^2}{T} \nabla T^4} \sim r^{15/11}.
\eea
From this characteristic cooling time we see that the smaller $r$ cool faster and the larger $r$ cool slower.
As a result, the center of the profile cools first and reaches a constant temperature.  As the outer regions start to cool, this region of constant temperature slowly expands in space while cooling off.  
As such, we will make the following approximation for the form of the cooling profile.
\bea
 T = T_{in} \qquad \qquad \qquad &\qquad& r < r_c\\
 T =  c \frac{\mpl^{4/11}}{r^{7/11}} = T_{in} \left ( \frac{r_c}{r} \right )^{7/11} &\qquad& r > r_c
\eea
We take there to be a cooling radius $r_c(t)$ inside of which there is a uniform sphere of constant temperature $T_{in}(t)$.  Outside of the cooling radius, the temperature is the same as it was pre-cooling.  Matching at the boundary relates 
\bea
\label{Eq: match}
T_{in} \sim \mpl^{4/11}/r_c^{7/11}.
\eea

We can find the functions $r_c(t)$ and $T_{in}(t)$ using conservation of energy.  The region of space inside of $r_c(t)$ is cooling at a rate
\bea
\label{Eq: cooling}
\frac{dE}{dt} \sim 4 \pi r_c^2 \frac{\lambda}{3} \nabla \rho \sim r_c T_{in}^3 \qquad \qquad \text{with} \qquad \qquad \frac{dE}{dt} \sim \frac{E}{t} \sim \frac{r_c^3 T_{in}^4}{t}.
\eea
Combining Eq.~\ref{Eq: match} with Eq.~\ref{Eq: cooling}, we find the time dependencies 
\bea
r_c(t)\sim \frac{t^{11/15}}{\mpl^{4/15}} \qquad T_{in}(t) \sim \frac{\mpl^{8/15}}{t^{7/15}}.
\label{Eq: cool}
\eea

\paragraph{Numerical Calculation}
How the profile cools is easy to solve numerically.  We numerically solve the conservation of energy equation, Eq.~\ref{Eq: solve},
with the initial condition
%\footnote{Having an initial temperature profile of $r^{-1/3}$ in the region $r<r_0$ has only a negligible effect in our result.} 
\bea\label{eq:boundarycondition}
&T(r,t_0)& =T_0 \qquad \qquad \qquad \qquad \qquad  r \leq r_0 \\
&T(r,t_0)& = T_0 \left ( \frac{r_0}{r} \right )^{7/11} \qquad \qquad \qquad r > r_0.
\eea
We add a small region of constant $T_0$ in the center of the initial profile to help deal with $r = 0$.  Our numerical results are insensitive to how one treats the $r \sim 0$ region.
%where $\beta=83 \lc_{\alpha} ^2 G_f/ g_*\r^{3/10}$ and the symbols have the same meaning as delineated in the Eq.~\ref{Eq:heatingprofile}.  

\begin{figure}[t]
    \centering
    \includegraphics[width=0.7 \textwidth]{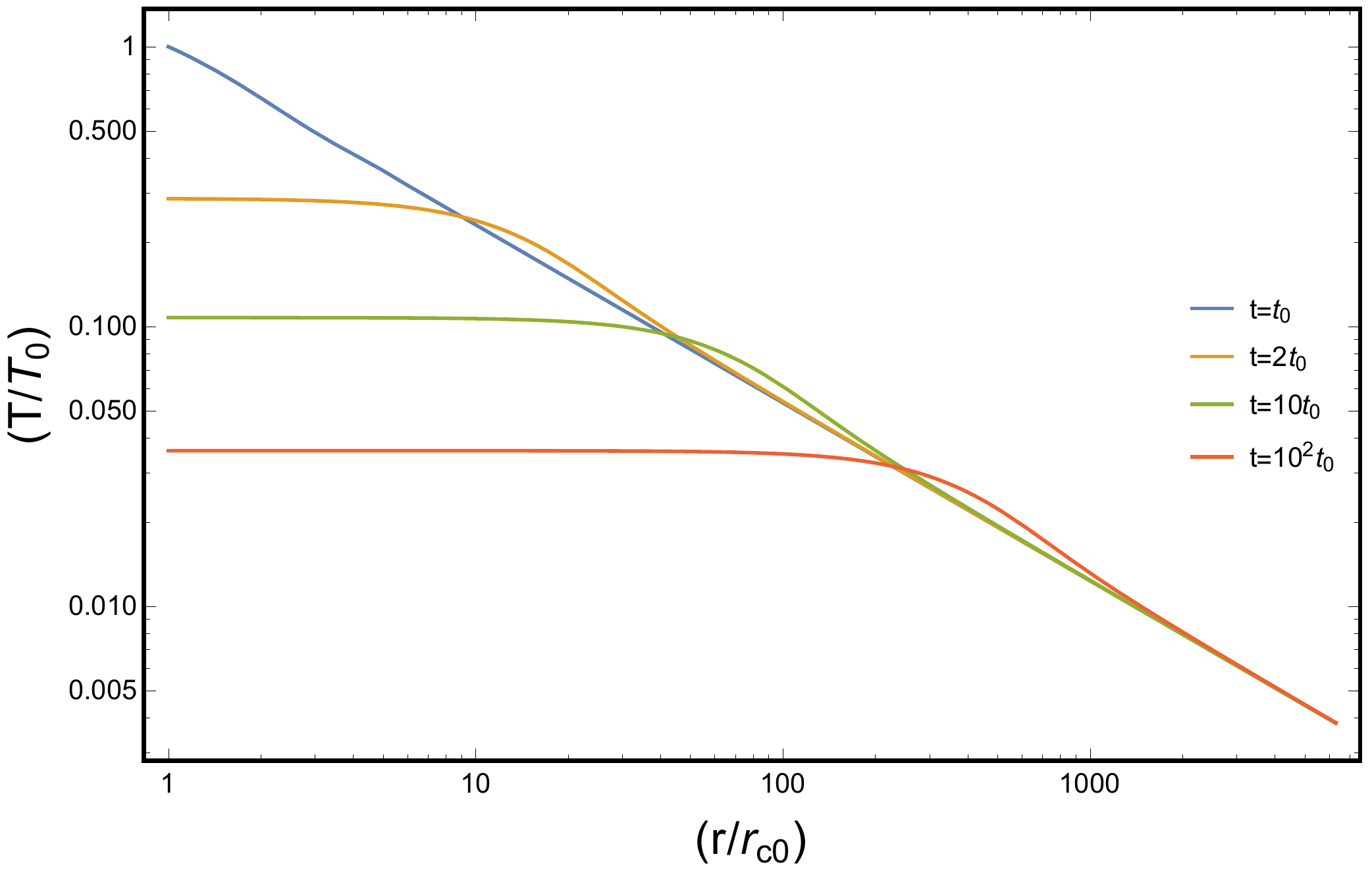}
    \caption{Temperature profile at different times where we have defined $r_{c0}=r_c(t_0)$. The initial profile follows Eq.~\ref{eq:boundarycondition}. As time progresses, the constant temperature region slowly expands without affecting the outside temperature significantly.}
    \label{fig: cooling}
\end{figure}

The results of the numerical simulation are shown in Fig.~\ref{fig: cooling}.  As is evident, the numerical solution satisfies our intuition about how cooling occurs.  Namely, there is a region of constant temperature that slowly expands as it cools off.  Meanwhile  the temperature outside this expanding sphere maintains its pre-cooling temperature.
The result of a numerical simulation is that the temperature of the plautau region falls as
\bea
T_{in} = T(r_0,t) = 0.87\l g_*(T)c_{\alpha}^2\r^{7/15} \frac{T_0^{22/15} r_0^{14/15}}{t^{7/15}}.
\label{Eq: coolnum}
\eea

Using the results of Eq.~\ref{Eq: coolnum} and Eq.~\ref{Eq:heatingprofile}, we find that the late-time cooling profile is
\begin{equation}
    \begin{split}
        \label{eq:numcooling}
        &T_{in}(t) =0.072\l c_{\alpha} ^{6/5}g_*(T)^{1/3} G_f^{2/15}\r \frac{\mpl^{8/15}}{t^{7/15}} . \\
          % &r_c(t)=31\l\frac{1}{c_{\alpha} ^{8/5} g_*(T)^{2/3}G_f^{1/15}}\r \frac{t^{11/15}}{\mpl^{4/15}}
    \end{split}
\end{equation}
This equation tells us how fast the plasma surrounding the black hole cools through the phase transition, which in turns determines the number of monopoles produced per black hole.  The longer the system remains in the symmetry restored phase, the fewer topological defects produced.

\section{Monopoles From Black Holes} \label{Sec: monopoles}

In this section, we apply our knowledge of the temperature profile around a black hole to calculate the number of monopoles produced per black hole.

\subsection{The Kibble-Zurek Mechanism}

The basic mechanism by which monopoles are produced is the Kibble-Zurek mechanism.  In this sub-section, we provide a brief review of the Kibble-Zurek mechanism as applied to thermal systems that are slowly cooling down~\cite{Murayama:2009nj}.
%Monopoles are the result of topologically non-trivial vacuum configurations and their production is all about how the universe relaxes to its minimum.

In many thermal systems, the high temperature limit involves a symmetry unbroken phase while the low temperature limit involves a symmetry broken phase.  Consider a hot plasma in the symmetry unbroken phase.
As a hot plasma cools, spontaneous symmetry breaking occurs through the condensation of a scalar order parameter $\Phi$.  When this happens, $\Phi$ randomly chooses an expectation value somewhere along its vacuum manifold.  
%In regions of space determined by the correlation length, the Goldstone bosons will choose the same value.  
Regions of space separated by distances longer than the correlation length $\xi$ will obtain different values of $\Phi$.  By random chance, these regions of space can accidentally form topologically non-trivial objects such as monopoles.  As a result, in a region of size $R$
\bea
\label{Eq: nm1}
N_M \sim \frac{R^3}{\xi^3}
\eea
monopoles are created.  It is not obvious what the value of the proportionality constant in front of Eq.~\ref{Eq: nm1} is, so for simplicity we will take it to be 1.

If the phase transition is first order, then over most of the parameter space a single bubble will nucleate and devour the entire space $R$ before a second bubble has a chance to form.  Depending on how this bubble interacts with the outside low temperature region, there will be at best $\mathcal{O}(1)$ number of monopoles created.  It is plausible that no monopoles will be created.  In some corners of parameter space, either because the bubble nucleation rate is highly sensitive to temperature or because the bubbles expand extremely slowly, it is possible that $\xi \lesssim R$, however this is not generically the case and thus we will instead consider second order phase transitions.

If the phase transition is second order, then we can have the situation where $\xi \ll R$ and a much larger number of monopoles can be created.  We can characterize how close a second order phase transition is to the critical temperature $T_{PT}$ with a parameter $\epsilon$
\bea
\epsilon = \frac{T - T_{PT}}{T_{PT}}.
\eea
It is sometimes also convenient to express this parameter in terms of the time to the phase transition $t$ and the characteristic cooling time of the phase transition $\tau_{\rm char}$
\bea \label{Eq: epsilon}
\epsilon \sim \frac{t}{\tau_{\rm char}}.
\eea
The correlation length and time are
\bea \label{Eq: correlations}
\xi \sim l_0 \epsilon^{-\nu} \qquad \tau \sim \tau_0 \epsilon^{-\mu}
\eea
where $l_0$ and $\tau_0$ are typical time and length scales in the problem.
The system freezes in place when the time to the phase transition $t$ is of order the relaxation time
\bea
t \sim \tau \qquad \tau_{\rm char} \epsilon \lesssim \tau_0 \epsilon^{-\mu},
\eea
where we have used Eq.~\ref{Eq: epsilon} and Eq.~\ref{Eq: correlations}. 
Solving for $\epsilon$ and plugging it back into Eq.~\ref{Eq: correlations}, we find
\bea \label{Eq: xichar}
\xi \sim l_0 \left ( \frac{\tau_{\rm char}}{\tau_0} \right )^{\frac{\nu}{1+\mu}}.
\eea

In the case of a plasma slowly cooling in time, the correlation length and time are governed by the mass of the radial mode.  Expanding the mass squared in a Taylor series around $T_{PT}$, we find
\bea
\xi, \tau \sim \frac{1}{m(T)} \sim \frac{1}{\sqrt{\frac{d m^2(T)}{dT} \left ( T - T_{PT} \right ) }} \sim \frac{1}{T_{PT} \sqrt{\epsilon}}.
\eea
Thus we are interested in the scenario where $l_0, \tau_0 \sim 1/T$ and $\nu = \mu = 1/2$ so that Eq.~\ref{Eq: xichar} gives
\bea
\label{Eq: xi}
\xi = \frac{\beta }{T} \left ( T \tau_{\rm char} \right )^{1/3},
\eea
where $\beta$ is a proportionality constant and without considering a specific model, it is impossible to specify the value of $\beta$.
In the case of a weakly coupled scalar whose thermal mass comes from a quartic coupling ($\lambda$), we have $\beta \sim 1/\lambda^{1/3}$ and so $\beta$ can potentially be parametrically larger than $\mathcal{O}(1)$ in the small $\lambda$ limit.  When estimating monopole production later on, we will take $\beta = 1$ with the understanding that there is some model dependence in the estimate.

\subsection{Kibble-Zurek Around Black Holes}

We are now in a position to estimate how many monopoles are produced per black hole.  The mechanism of monopole production is that each black hole heats up the surrounding plasma to a temperature profile shown in Eq.~\ref{Eq: heat}.  After the black hole has evaporated, it cools down with a characteristic time scale shown in Eq.~\ref{Eq: cool}.  As the plasma cools past the phase transition temperature, the Kibble-Zurek mechanism generates some number of monopoles.
As mentioned before, if the phase transition was first order, then there are generically at most $\mathcal{O}(1)$ and possibly no monopoles produced per black hole.

The more interesting case is if the phase transition was second order.  Let us take the phase transition to occur at a scale $T_{PT}$.  Using Eq.~\ref{Eq: heat}, the radius of the region with $T > T_{PT}$ is
\bea\label{Eq:RPT}
R_{PT} \sim \frac{\mpl^{4/7}}{T_{PT}^{11/7}}.
\eea
Meanwhile, the characteristic timescale associated with cooling can be read off of Eq.~\ref{Eq: cool}
\bea\label{Eq:quenchtime}
\tau_{\rm char} \sim \frac{\mpl^{8/7}}{T_{PT}^{15/7}}.
\eea
Using Eq.~\ref{Eq: xi}, we find that the number of monopoles produced per black hole is
\bea\label{Eq:monopoleperBH}
N_m \sim \frac{R_{PT}^3}{\xi^3} \sim \left ( \frac{\mpl}{T_{PT}} \right )^{4/7} .
\eea
Thus a significant number of monopoles can be potentially produced per black hole.

\section{Bounds on Reheating from Black Holes} \label{Sec: reheat}

In this section, we place reheating bounds on the scenario where the decay of a population of black holes with the same mass reheats the universe.  Because each black hole can produce many monopoles, monopoles have the possibility of overclosing the universe.

We estimate the bounds on the reheating temperature in two steps.  We first omit all $\mathcal{O}(1)$ numbers in order to emphasize the scaling behavior.  Afterwards, we redo the estimate using all of the $\mathcal{O}(1)$ numbers.

If black holes are responsible for reheating the universe, the black holes decay when  
\bea
H^2 \sim \frac{\trh^4}{\mpl^2} \sim \frac{1}{t_{BH}^2} \sim \frac{\mpl^8}{M_{BH}^6}.
\eea
Using this, the number density of black holes over the number density of photons is given by
\bea
Y_{BH} = \frac{n_{BH}}{s} \sim \left ( \frac{\trh}{\mpl} \right )^{5/3}.
\eea
Bounds on overclosing the universe can be obtained by requiring that the energy density in monopoles, $M N_m Y_{BH}$ is smaller than roughly five times the energy density in baryons~\cite{Planck:2018vyg}, $m_B Y_B$.  Taking the mass of the monopole $M$ to be 100 $T_{PT}$ and using Eq.~\ref{Eq:monopoleperBH} for $N_m$, we find the bound on the reheat temperature to be
\bea
T_{RH} \lesssim 500 \, \text{GeV} \, \left ( \frac{10^{15} \, \text{GeV}}{T_{PT}} \right )^{9/35} \qquad \qquad \text{$\mathcal{O}(1)$ Estimate}.
\eea
This shows that we can expect a very strong bound on the reheat temperature in these scenarios.

Now we produce a more refined estimate by keeping all $\mathcal{O}(1)$ numbers. Using the results of Sec.~\ref{Sec: temp}, Eq.~\ref{Eq:RPT} and Eq.~\ref{Eq:quenchtime} with  $\mathcal{O}(1)$ numbers become
\begin{equation}\label{Eq:RPTO1}
    \begin{split}
        &R_{PT}=0.07 \l{\frac{c_{\alpha} ^2 G_f}{g_*}}\r^{1/7}\l\frac{\mpl^{4/7}}{T_{PT}^{11/7}}\r\\
        &\tau_{\rm char}=0.008\l c_{\alpha} ^{18} g_*^5 G_f^2\r^{1/7}\l\frac{\mpl^{8/7}}{T_{PT}^{15/7}}\r .
    \end{split}
\end{equation}
Using these equations, we can calculate the numerical coefficient in Eq.~\ref{Eq:monopoleperBH} and find
\bea \label{Eq: NMO1}
N_m =0.044\l\frac{G_f}{c_{\alpha} ^{12} g_*^8}\r^{1/7} \l  \frac{\mpl}{T_{PT}} \r ^{4/7} .
\eea
For the last step, we assume that black holes decay instantaneously when $H = 1/t_{BH}$.
Using the fiducial values $G_f=3.8$, $g_*(T_{PT})=108$, $c_{\alpha}=1/10$ and $M=25\, T_{PT}$ at the unification scale, we arrive at
\bea\label{Eq:FinalReheat}
T_{RH} \lesssim 672 \, \text{GeV} \, \left ( \frac{10^{15} \, \text{GeV}}{T_{PT}} \right )^{9/35}.
\eea
Thus we see that having a large reheat temperature runs the risk of over-producing monopoles.

%\AH{What are the various assumptions going into all of this?}
%We have assumed the uncomplicated scenario in which the universe was filled with identical black holes which evaporated at the same time. We also have assumed that the black holes are separated by large enough distances that evaporation of one is completely unaffected by the others.

\section{Realm of validity}
\label{Sec: approx}

In this section, we discuss the various approximations that go into our result and the limitations placed on our result by these approximations.

\subsection{Model dependent factors}

Many of our results depend to some degree on the model dependent factors $c_{\alpha}$, $\beta$, $g_\star(T)$ and $G_f$.  
Of these, it is expected that $g_\star$ and $G_f$ change by at most $\mathcal{O}(1)$ and thus do not change the final results by much.
On the other hand, it is possible for $c_{\alpha}$ and $\beta$ to change by more than an order of magnitude and can thus change the final result more significantly.

As mentioned before, a major source of uncertainty comes from the the correlation length at criticality, $\xi$.  $\beta$ appears as a proportionality constant in Eq.~\ref{Eq: xi} and its value depends on the Grand Unified Theory under consideration.  The expectation is that $\beta \gtrsim \mathcal{O}(1)$.  
While the exact value of $\beta$ is unknown, the parametric dependence of our final reheat temperature on $\beta$ is easily calculated. A larger correlation length decreases the abundance of magnetic monopoles which makes the bound on $T_{RH}$ weaker.  It is easy to verify that the monopole abundance decreases by an amount $1/\beta^3$ which weakens the bound on the reheat temperature by $\beta^{9/5}$ in  Eq.~\ref{Eq:FinalReheat}.
\bea
T_{RH} \lesssim 672 \, \text{GeV} \, \beta^{\frac{9}{5}} \left ( \frac{10^{15} \, \text{GeV}}{T_{PT}} \right )^{9/35}.
\eea

The other important model dependent quantity is $c_{\alpha}$, which characterizes the typical size of the scattering cross sections.  $c_{\alpha}$ is important because scattering is responsible for the diffusion of energy, which affects the profile.  We have defined the typical interaction cross section $\sigma$ as
\bea
\sigma\equiv \frac{c_{\alpha}^2}{T^2}
\eea
in the high energy limit.  
Larger $c_{\alpha}$ give shorter diffusion lengths and so that energy diffuses out more slowly.
As a result, in equilibrium the temperature distribution changes more gradually and there is a larger volume at higher temperature.  But larger $c_{\alpha}$ (slower diffusion) also hinders cooling which makes the correlation length at freeze out at larger values. Between these two effects, the effect on the correlation length is stronger as can be seen in Eq.~\ref{Eq: NMO1}.
As a result of this, the final bound on the reheat temperature scales as
\bea
T_{RH} \lesssim 672 \, \text{GeV} \, \beta^{9/5}\l\frac{g_*(T_{PT})}{108}\r^{11/14} \l\frac{G_f}{3.8}\r^{4/35}\l 10c_{\alpha}\r^{36/35} \left ( \frac{10^{15} \, \text{GeV}}{T_{PT}} \right )^{9/35}.
\eea

As will be shown later, a critical assumption of our derivation is that the thermal bath is in equilibrium with the evaporating black hole.  The validity of this assumption is $c_{\alpha}$ dependent and may be where the uncertainty in $c_{\alpha}$ is most important.
%about thermal equilibrium and success of Kibble-Zurek Mechanism in producing magnetic monopoles by evaporating black holes hinges on detail of interaction strength which we will discuss next.

\subsection{Hierarchy of length scales}

In our previous derivation, it was tacitly assumed that $R_{PT} > \xi$ and $\xi > r_m$, where $r_m$ is the size of the monopole.  In this subsection, we discuss the validity of these assumptions.
Our work is based on the premise that an evaporating BH will heat up a large volume of the surrounding plasma, where the phase transition can take place and produce topological defects. This assumes that the size of the region that attains temperatures above $T_{PT}$ is larger than the correlation length, namely $R_{PT}>\xi$. 
In the limit that $R_{PT} < \xi$, then like first order phase transitions, either $\mathcal{O}(1)$ or zero monopoles will be produced and our estimate would need to be modified. 

The second inequality comes from the fact that the monopole is an extended object with a characteristic length scale $r_m$.  We assumed that one monopole was produced per volume $\xi^3$, an assumption only valid if $\xi>r_m$.  In the limit $\xi < r_m$ there is one monopole produced per $r_m^3$ instead of $\xi^3$, and the estimate must be modified.

The temperature dependence of $R_{PT}$ and $\xi$ can be seen in Eq.~\ref{Eq: xi} and Eq.~\ref{Eq:RPT}, namely $R_{PT}\sim T_{PT}^{-11/7}$ and $\xi\sim T_{PT}^{-29/21}$.  Meanwhile, the radius of the monopole scales with its mass as $r_m\sim T_{PT}^{-1}$. 
Putting in the relevant pre-factors, we find
\begin{equation}
    R_{PT}=0.07\l{\frac{c_{\alpha} ^2 G_f }{g_* }}\r^{1/7}\l\frac{\mpl^{4/7}}{T_{PT}^{11/7}}\r \qquad \xi = 0.20\l c_{\alpha}^{18}g_*^5 G_f^2\r^{1/21} \l\frac{ \mpl^{8/21}}{T_{PT}^{29/21}}\r .
\end{equation}
From this, we see that $R_{PT}$ falls off fastest with increasing phase transition temperature while $\xi$ and $r_m$ decrease more slowly.  
Depending on the value of $g_*(T)$ and $c_{\alpha}$,  one of either $R_{PT} > \xi$ or $\xi > r_m$ is more important. Combining the two, we find that as long as 
\bea
    T_{PT}< 
\begin{cases}
    1.37\times 10^{16} \, \text{GeV} \l\frac{108}{g_*(T_{PT})}\r^2 \l\frac{G_f}{3.8}\r^{1/4}\l\frac{0.076}{c_{\alpha}}\r^{3},& \text{if } g_*(T)c_{\alpha}^2(T)\geq 0.63\\
    1.37\times 10^{16} \, \text{GeV} \l\frac{g_*(T_{PT})}{108}\r^{5/8} \l\frac{G_f}{3.8}\r^{1/4}\l \frac{c_{\alpha}}{0.076} \r^{9/4},              & \text{otherwise}
\end{cases}
\eea
our assumptions ($R_{PT} > \xi > r_m$) are valid. For our fiducial parameters, $g_*(T)c_{\alpha}^2=1.08$, the conditions are safely satisfied for $T_{PT}< 6.1\times 10^{15} \, \text{GeV}.$

\subsection{Thermal equilibrium}
% How close to the BH is temperature profile valid?

% How high can $T_{PT}$ be and still work ok?

% Gradient smaller than scattering length.

% scattering length $<r$\\

Throughout our work, we have assumed thermal equilibrium.  There are several areas where the approximation of thermal equilibrium break down.  For example if the temperature is changing on length scales shorter than the scattering length, then it is clear that thermal equilibrium is breaking down.  In the context of monopole production, the most important and constraining assumption that was made was the assumption that that the energy emitted by the black hole is in thermal equilibrium with the surrounding plasma. 

The Hawking radiation emitted by the black hole only reaches thermal equilibrium when it has lost all of its energy.
All of the Hawking radiation eventually scatters and loses its energy, so for a large enough radius, we expect the thermal equilibrium condition to hold.  We will thus assume the equilibrium configuration to hold until some critical radius $r_{th}$ inside of which the system is not in thermal equilibrium.
The equilibrium temperature distribution can be found by requiring that the power passing through the temperature distribution match the power emitted by the black hole, $r T(r)^3 \sim T_{BH}^2$.

As the radiation emitted by the black hole with energy $E$ passes through the plasma, it loses energy through its interaction as described in Eq.~\ref{Eq: rad1}
 \begin{equation}
     \partial_r E = -n\, \sigma \, E \sim T^2.
 \end{equation}
While the number density of the thermal bath scales as $n \sim T^3$, the cross section of a high energy particle scales as $\sigma \sim 1/s \sim 1/T E$ resulting in a constant energy loss of order $T^2$ when moving through the plasma.
Hawking radiation loses all of its energy after traveling a distance
\bea
r_{th} \sim \frac{T_{BH}}{T(r_{th})^2} \qquad \qquad r_{th} \sim \frac{1}{T_{BH}} .
\eea

Since $\partial_r E \sim -T^2$, high energy particles travel longer before they they thermalize. Averaging over the energies in the relativistic limit, we get the following $\mathcal{O}(1)$ factors
\bea\label{Eq: rth}
%&r_{th} =r_{BH}+\l 54 \pi ^2 c_{\alpha} ^4 g^2 \zeta (3)^2+576 \pi ^5 c_{\alpha} ^2 g \zeta (3)+2560 \pi ^8\r /(9 c_{\alpha} ^6 g^3  \zeta (3)^3 \, T_{BH}) \\
&r_{th}\approx \frac{ 1.3\times 10^6}{T_{BH}} \l g_*(T)c_{\alpha}^2\r^{-3} \qquad T(r_{th} ) \approx 0.004 \, T_{BH} \,  g_*(T)c_{\alpha}^2  .
\eea
From this, we can see that unless one is interested in temperatures close to $T_{BH}$, the assumption of thermal equilibrium is good.

We can finally use Eq.~\ref{Eq: rth} to find a maximum for $T_{PT}$.  Given that we assumed that the system was in equilibrium when $T_{PT}$ froze out, we can require that $R_{PT} > r_{th}$ when the $T_{PT}$ froze out.  Imposing this requirement gives
\bea
T_{PT} \lesssim 1.94\times 10^{15}\l\frac{g_*(T)c_{\alpha}^2}{1.08}\r \, \text{GeV}.
\eea
which is satisfied for our choice of $T_{PT}$.

\subsection{Evaporation of Black Holes}

If thermal equilibrium was maintained all the way to the surface of the black hole, then the evaporation of the black hole would be greatly affected.  In thermal equilibrium, radiation into and out of the black hole would roughly balance and only temperature gradients give rise to an outflow of energy.  
As was demonstrated in the previous subsection, the area right outside of the black hole is not in thermal equilibrium with the black hole itself.  As a result, its temperature is lower than the black hole temperature, see e.g. Eq.~\ref{Eq: rth}.  Thus the energy falling into the black hole, $r_{BH}^2 T(r_{th})^4$,  is subdominant to the energy being emitted by the black hole, $r_{BH}^2 T_{BH}^4$.  As such, we can treat the black hole as evaporating the same as it would in vacuum.

\subsection{Reaching thermal equilibrium}

When the black holes first start evaporating, they are not in equilibrium with the external radiation.  In this subsection, we estimate the time it takes for the plasma surrounding the black hole to reach an equilibrium state.

When the thermal profile is close to the equilibrium profile, the black hole's ability to heat the surrounding plasma is limited by diffusion.  Following the argument in Sec.~\ref{Sec:anaheating}, we find
\bea
r_{d}^2 \sim \lambda t \qquad r_d \sim \frac{t^{11/15}}{\MP^{4/15}}.
\eea
As long as the black hole is depositing energy in a radius smaller than $r_d$, then in a time $t$, the region of space with $r < r_d$ will be in equilibrium with the black hole.

Initially, the plasma surrounding the black hole has an initial temperature $T_{RH}$.  The Hawking radiation emitted by the black hole loses all of its energy after a distance, see Eq.~\ref{Eq: rad1},
\bea
r_i \sim \frac{T_{BH}}{T_{RH}^2}.
\eea
This is the radius at which the black hole first starts to deposit its energy.  As the temperature rises, it deposits energy closer and closer to the black hole.  As long as $r_i < r_d(t)$, the black hole will have reached equilibrium with the plasma immediately around it.  Equilbrium is first reached when $r_i \sim r_d(t_i)$ giving
\bea
t_i \sim \frac{T_{BH}^{15/11} \MP^{4/11}}{T_{RH}^{30/11}}.
\eea

When black holes are reheating the universe, $t_i$ and $T_{RH}$ are set by the lifetime of the black holes and Hubble.  Comparing the two timescales, we find
\bea\label{Eq:equitime}
t_i \sim t_{BH} \left ( \frac{T_{BH}}{\MP} \right )^{3/11} < t_{BH}
\eea
showing that the plasma around the black hole does indeed have time to reach equilibrium with the black hole. For our choice of parameters, the $O(1)$ number in Eq.~\ref{Eq:equitime} is $0.9$, which validates our assumption that the black hole reaches thermal equilibrium with its surrounding.

\section{Conclusion}\label{Sec: conclusion}

In this article, we explored the evaporation of black holes in the early universe and showed how they heat up the surrounding plasma.  This plasma can reach temperatures much larger than the ambient temperature and can have effects more significant than that of the black hole itself.  As an example, while monopole production by black hole evaporation is negligible,  monopole production by the surrounding plasma can be very significant.  This efficient mechanism of monopole production can be significant enough that it can easily overclose the universe if the reheat temperature is larger than $T_{RH} \gtrsim 500$ GeV.  Such a low reheat temperature motivates black hole centric mechanisms of baryogenesis.

The evaporation of black holes in the early universe is an intriguing possibility.  We have only listed a single scenario where the plasma surrounding the black hole has a significant effect.  It would be interesting if there are other situations where this plasma is important.

\acknowledgments{
We thank Prateek Agrawal for helpful comments on the draft.  This research was supported in part by the NSF under Grant No. PHY-1914480 and by the Maryland Center for Fundamental Physics (MCFP).
}

\appendix

%\section{Reheating Bounds}

\bibliographystyle{JHEP}
\bibliography{references.bib}

\end{document}